\documentclass[%
 aip,
 sd,%
 amsmath,amssymb,
 reprint,%
]{revtex4-1}

\usepackage{graphicx}
\usepackage{dcolumn}
\usepackage{bm}
\usepackage{array}
\usepackage[table]{xcolor}

\begin{document}

\preprint{AIP/123-QED}

\title[]{The new high field photoexcitation muon spectrometer\\at the ISIS pulsed neutron and muon source}

\author{K. Yokoyama}
\email{koji.yokoyama@stfc.ac.uk}
\affiliation{ 
School of Physics and Astronomy, Queen Mary University of London, Mile End, London, E1 4NS, United Kingdom
}%
\affiliation{%
ISIS, STFC Rutherford Appleton Laboratory, Didcot, OX11 0QX, United Kingdom
}%

\author{J. S. Lord}
\affiliation{%
ISIS, STFC Rutherford Appleton Laboratory, Didcot, OX11 0QX, United Kingdom
}%

\author{P. Murahari}
\affiliation{ 
School of Physics and Astronomy, Queen Mary University of London, Mile End, London, E1 4NS, United Kingdom
}%

\author{K. Wang}
\affiliation{ 
School of Physics and Astronomy, Queen Mary University of London, Mile End, London, E1 4NS, United Kingdom
}%
\affiliation{%
College of Physical Sciences, Sichuan University, P. R. China
}%

\author{D. J. Dunstan}
\affiliation{ 
School of Physics and Astronomy, Queen Mary University of London, Mile End, London, E1 4NS, United Kingdom
}%

\author{S. P. Waller}
\author{D. J. McPhail}
\author{A. D. Hillier}
\affiliation{%
ISIS, STFC Rutherford Appleton Laboratory, Didcot, OX11 0QX, United Kingdom
}%

\author{J. Henson}
\author{M. R. Harper}
\affiliation{%
Litron Lasers Ltd., Rugby, CV21 1PB, United Kingdom
}%

\author{P. Heathcote}
\affiliation{ 
School of Biological and Chemical Sciences, Queen Mary University of London, Mile End, London, E1 4NS, United Kingdom
}%

\author{A. J. Drew}
\email{a.j.drew@qmul.ac.uk}
\affiliation{ 
School of Physics and Astronomy, Queen Mary University of London, Mile End, London, E1 4NS, United Kingdom
}%
\affiliation{%
ISIS, STFC Rutherford Appleton Laboratory, Didcot, OX11 0QX, United Kingdom
}%
\affiliation{%
College of Physical Sciences, Sichuan University, P. R. China
}%

\date{\today}

\begin{abstract}
A high power pulsed laser system has been installed on the high magnetic field muon spectrometer (HiFi) at the ISIS pulsed neutron and muon source, situated at the STFC Rutherford Appleton Laboratory in the UK. The upgrade enables one to perform light-pump muon-probe experiments under a high magnetic field, which opens new applications of muon spin spectroscopy. In this report we give an overview of the principle of the HiFi Laser system, and describe the newly developed techniques and devices that enable precisely controlled photoexcitation of samples in the muon instrument. A demonstration experiment illustrates the potential of this unique combination of the photoexcited system and avoided level crossing technique.
\end{abstract}

\maketitle

\section{introduction}
\label{sec:Introduction}
Muon spin spectroscopy (collectively known as $\mu$SR, corresponding to muon spin relaxation/rotation/resonance) has shown itself to be a powerful probe of material properties, with major studies directed towards understanding semiconductors, magnetic materials, superconductors, organic materials, and many other systems. \cite{Schenk, Nuccio, Wang, McKenzie} Spin-polarized positively charged (anti-)muons with an energy of 4.2 MeV  are generated in a proton accelerator facility and implanted in bulk materials with the distribution thermalizing over several hundred $\mu$m. The muons then decay with a lifetime of 2.2 $\mu$s and emit positrons preferentially in the muon spin direction at time of decay, which is then subsequently detected. Since the number of implanted muons is small for the volume, the interaction between them is totally negligible. Therefore $\mu$SR is a unique and ideal real-space probe to investigate local magnetic fields in materials. In some circumstances implanted muons capture an electron to form a muonium (Mu = $\mu$$^+$ + $e$$^-$), which is an analogue of hydrogen atom, and thus is an important probe of the behavior of hydrogen in various material systems such as semiconductors and insulators. \cite{Patterson, CoxRev, Chow, CoxOxide}

There are two types of muon sources with different time structures: quasi-continuous and pulsed. The continuous sources, such as PSI in Switzerland and TRIUMF in Canada, provide a quasi-continuous beam of muons with weak modulation at the RF frequency of the accelerator, typically 50 MHz. In the pulsed sources, ISIS in the UK and J-PARC MLF in Japan, muons come in intense bunches (70 ns FWHM in ISIS) at a repetition rate of a few tens of Hz. There are advantages and disadvantages for both cases, and one of the advantages of a pulsed source is that it is well matched to performing experiments with pulsed stimulations, such as RF, electric field, and light. \cite{Cottrell, Giblin, Storchak, Yokoyama} In addition to being able to perform traditional pump-probe experiments, having a pulsed stimulation enables one to achieve a larger stimulation by virtue of high peak intensity.

Among those different types of stimulation, photoexcited $\mu$SR studies (termed ``photo-$\mu$SR'') have attracted a persistent interest in the $\mu$SR community, and been applied to various experimental systems. \cite{Yokoyama} One of the major applications of photo-$\mu$SR in condensed matter physics is focused on a variety of semiconductors. The light effect is manifested as an additional depolarization in the $\mu$SR time spectrum, which gives information not only on the spin exchange interaction between the Mu centers and injected excess electrons, but also on the charge- and site-exchange dynamics of Mu. \cite{KadonoPRL, KadonoPRB, Prokscha, Fan, FanGe, YokoyamaGaAs, Shimomura} Some of these experiments have utilized flashlamps or light bulbs as a convenient light source, \cite{KadonoPRL, KadonoPRB, Fan, FanGe} although quantifying their optical parameters is usually difficult. On the other hand, although costing more and being more complex, lasers surpass flashlamps or light bulbs in every aspect: tunable wavelength, linewidth, pulse width, beam profile and pointing, etc. One can measure and control these optical parameters accurately, which is crucial to characterize the light-induced effect. Currently there are two spectrometers in the world, that are capable of performing pulsed photo-$\mu$SR experiments with a laser: ARGUS and HiFi at ISIS. In this report, we introduce the recently commissioned HiFi photo-$\mu$SR spectrometer, which provides a new experimental environment by combining a high power laser system and a high magnetic field.

The ISIS pulsed muon facility at the STFC Rutherford Appleton Laboratory is currently home to five muon instruments, each having unique capabilities. \cite{King} One of the spectrometers, HiFi, can apply a magnetic field on a sample up to 5 T, and is optimized for time-differential muon spin relaxation studies. \cite{Lord} Such high field is especially useful in the avoided level-crossing (ALC) resonance technique, which probes the Mu energy states by scanning and finding a field where two Mu levels cross over. \cite{Nuccio, Wang, McKenzie, KieflEstle} At the specific field the muon spin is depolarized either because 1) the muon polarization is transferred to neighboring nuclear spins due to quantum mechanically coupled states (so-called $\Delta_0$ transition), or 2) the muon spin flips due to its precession around the symmetry axis of Mu (this $\Delta_1$ transition is usually only observable in solids). The ALC technique has been intensively applied on Mu centers in various semiconductors to study the electronic structure of Mu and its local hyperfine (HF) environment. \cite{KieflEstle} Excess carriers from photoinjection or other sources induce electron spin relaxation of Mu, which results in modification of the ALC spectrum. A better understanding of the carrier-Mu interaction should be obtained using the laser excitation, which can generate high excess carrier density and control it accurately.

Another important ALC application is to study muoniated molecular radicals either in solution or the solid state, where a generated Mu forms a covalent bond and creates radical states of the target molecule. The delocalized electronic wave function, although restricted in proximity to the parent muon due to the Coulomb attraction, spreads over the molecule and mediates a HF interaction between the muon and nearby nuclei. Depending on the attached site, the ALC spectra can have a different amplitude, central field, width, and shape. This method has been applied to many hydrocarbon molecules, \cite{McKenzie} and recently found its application in organic semiconductor materials in the solid state. \cite{Nuccio, Wang, Nuccio13, Schulz11}  Based on the unique capability of observing the local electronic/nuclear spin environment in molecules, it is natural to extend its application to observe behavior of molecules excited by light, a crucial subject in many biological and chemical processes. Photoexcitation of an electron from the HOMO to LUMO state results in a change in the electronic distribution of a molecule. Therefore, upon attachment to a photoexcited molecule, the Mu should see a different electronic environment compared with its ground state. It is predicted that this change in electronic environment will be manifested as a change in the ALC spectra, which should carry information on the excited electronic structure and dynamics, by virtue of the positional sensitivity brought about by the different Mu adducts and thus unique ALC resonance positions. 

In addition to these future photo-$\mu$SR experiments with the ALC technique, the upgraded instrument is suitable for general-purpose photo-$\mu$SR experiments. For instance, high spectral intensity enables one to perform optical spectroscopy of Mu in materials. Mu centers can form defect levels within a band gap in many semiconductors. \cite{CoxRev} It may be possible to photoionize the Mu centers and identify their energy levels optically. \cite{Shimomura2} An injection seeded single frequency laser can extend the method to gas phase to study photoexcited molecules. \cite{Bakule} With a monochromatic and linearly polarized light source, one can easily generate and manipulate circularly polarized light, which is especially useful to study optically induced electronic/nuclear spin and its interactions with Mu. \cite{YokoyamaGaAs} Considering optical spectroscopy has been a highly versatile method in many science subjects and developed a rich diversity of approaches and techniques, we should find the great potential also in photo-$\mu$SR.

In Sec. \ref{sec:LaserSystem} we first describe the laser system and its performance. We then illustrate the laser cabin and beam transport system in Sec. \ref{sec:LaserBeamTransport}. Sec. \ref{sec:ExperimentalTechniques} focuses on various technical details of photo-$\mu$SR experiments, including the sample environments. Then finally Sec. \ref{sec:Silicon} shows a commissioning experiment on silicon demonstrating a principle of the unique combination of the photo-$\mu$SR and ALC techniques.

\section{Laser system}
\label{sec:LaserSystem}
To perform the photo-$\mu$SR experiments envisaged above, the laser system should provide: 1) short pulses comparable to the timescale of $\mu$SR, 2) high peak intensity, 3) wide wavelength tunability, 4) narrow linewidth, and 5) long-term stability. Based on these requirements we have chosen a flashlamp-pumped Q-switched nanosecond Nd:YAG laser (YAG) and a 355-nm pumped optical parametric oscillator (OPO). Experiments requiring a high-energy pulse can utilize the YAG fundamental and harmonics, whereas experiments requiring wavelength tuning can use the OPO. Specifications of the laser system are summarized in TABLE \ref{table:LaserSpecs}.

\begin{table*}
\centering
\caption{Laser system specifications}
\label{table:LaserSpecs}

\begin{tabular}{|>{\centering\arraybackslash}p{23mm}|>{\centering\arraybackslash}p{10mm}|>{\centering\arraybackslash}p{10mm}|>{\centering\arraybackslash}p{9mm}|>{\centering\arraybackslash}p{9mm}|>{\centering\arraybackslash}p{9mm}|>{\centering\arraybackslash}p{36mm}|>{\centering\arraybackslash}p{32mm}|>{\centering\arraybackslash}p{25mm}|}
\hline
\rowcolor{lightgray} \multicolumn{9}{|c|}{Nd:YAG laser}\\
\hline
Repetition rate& \multicolumn{5}{|c|}{Maximum pulse energy [mJ]} & Pulse duration (FWHM) & Linewidth (FWHM) & Beam intensity \\

[Hz] & 1064 & 532 & 355 & 266 & 212 & for 1064 nm [ns] & for 1064 nm [cm$^{-1}$] & profile\\
\hline \hline
25 & 2100 & 1200 & 415 & 160 & 51 & 16 & $\approx$1 \footnotemark[1] & multi-mode\\
\hline 
\end{tabular}
\footnotetext[1] {can be narrowed to 0.003 cm$^{-1}$ with injection seeding (not available yet).}

\begin{tabular}{|>{\centering\arraybackslash}p{16mm}|>{\centering\arraybackslash}p{17mm}|>{\centering\arraybackslash}p{15mm}|>{\centering\arraybackslash}p{15mm}|>{\centering\arraybackslash}p{36mm}|>{\centering\arraybackslash}p{31mm}|>{\centering\arraybackslash}p{31mm}|}
\hline
\rowcolor{lightgray} \multicolumn{7}{|c|}{OPO}\\
\hline
\multicolumn{2}{|c|}{Tuning range [nm]} & \multicolumn{2}{|c|}{Pulse energy at peak [mJ]} & Pulse duration (FWHM) & Linewidth (FWHM) & Energy stability \\
signal & idler & 440 nm & 941 nm & [ns] & [cm$^{-1}$] & in 440 nm signal [\%] \\
\hline \hline
410 $\sim$ 710 & 710 $\sim$ 2400 & 20  \footnotemark[2] & 23  \footnotemark[2] & 11 & \verb|<|5 & $\pm$5 \\
\hline 
\end{tabular}
\footnotetext[2] {pumped with 240 mJ of THG pulse.}

\end{table*}

\subsection{Nd:YAG laser system}
\label{sec:NdYAG laser system}

\begin{figure}
\includegraphics{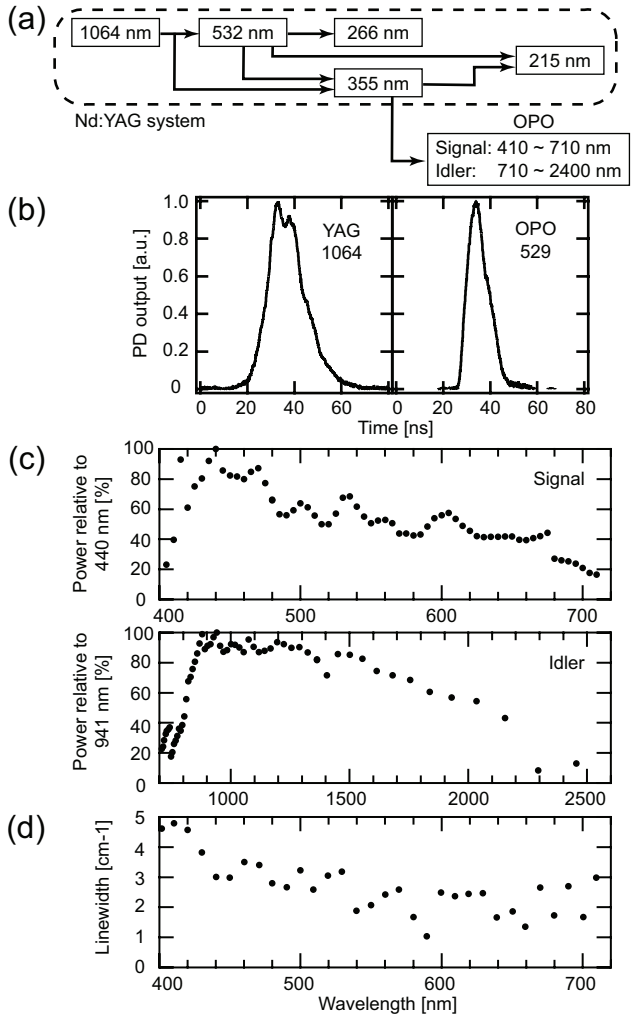}
\caption{\label{fig:Laser} (a) Wavelength conversion diagram of the HiFi Laser system. (b) temporal profile of the YAG fundamental (1064 nm) and OPO signal (529 nm) pulse. (c) Output spectrum for OPO signal (top) and idler (bottom). Uneven efficiency over the entire tuning range is a typical feature of the type-II BBO OPO. (d) Linewidth (in cm$^{-1}$) of OPO signal versus the wavelength.}
\end{figure}

The laser system is custom-made by Litron Lasers Ltd. for a high-energy pulse and automated beam path configuration. Fig. \ref{fig:Laser}(a) shows a diagram of the wavelength conversion. The two-stage YAG amplification stages with 15 mm diameter rods generate a 1064-nm laser pulse with energies up to 2100 mJ at 25 Hz with pulse width of $\approx$16 ns (FWHM). As shown in Fig. \ref{fig:Laser}(b), its temporal profile has a few peaks due to multiple longitudinal modes. As a future development, the YAG oscillator can be injection seeded for the single frequency operation. The spatial mode has a round flat-top multi-mode in the near-field, and deforms to an elliptical shape in the far-field (8 m).

The laser system can generate all YAG harmonic wavelengths up to the 5$^{th}$ harmonic.\cite{Note5th} With a computer command, the motorized crystal stages and mirror assemblies automatically alter the internal configuration of the laser to generate the requested harmonic. This is especially useful when a laser operator needs to quickly alter the output wavelength from one experiment to another. Their position can be reproduced accurately so that they require little adjustment of the phase-matching angle of crystals and the mirror position. 

\subsection{Optical parametric oscillator}
\label{sec:Optical parametric oscillator}
The 3$^{rd}$ harmonic pumps the OPO and generates a signal and idler beam, which cover wavelength in 405 - 710 nm and 710 - 2400 nm ranges respectively. \cite{NotePPE} Fig. \ref{fig:Laser}(c) shows the output spectrum for the signal and idler plotted relative to their maximum output energy. The use of type-II phase matching within a BBO (Beta Barium Borate) crystal ensures output at all wavelengths including at degeneracy.

Although the output power increases linearly with the pump energy, the pump is typically set below 240 mJ/pulse to ensure that it is well below the damage threshold and the BBO crystal can be used for an extended period. This pump energy generates pulse energies of 20 and 23 mJ/pulse at 440 and 941 nm respectively. The pulse width of OPO output is typically $\approx$11 ns (FWHM) as shown in Fig. \ref{fig:Laser}(b). The linewidth of OPO signal is less than 5 cm$^{\verb|-|1}$ over the entire tuning range (see Fig. \ref{fig:Laser}(d)), which should be enough for most applications in condensed matter physics. For type-II mixing, the relatively large birefringent walk-off combined with multiple passes through the crystal (as in a standard OPO cavity) causes a large beam divergence in the vertical direction. The mode, however, can be corrected by telescoping with a few spherical and cylindrical lenses. The collimated beam can then be transported to the HiFi instrument.

\subsection{Automation}
Automation of an experiment is often crucial to run a series of measurements with changing experimental parameters. This is typical in a $\mu$SR experiment, where one needs to run it continuously during the limited period of beam time. Currently the script running on the in-house developed sample environment control system (``SECI'') at ISIS, which controls muon data taking, can change three parameters on the laser: OPO wavelength, pulse energy, and the pump-probe pulse timing. The OPO wavelength can be changed by rotating the BBO crystal on a motorized rotation stage, which changes the phase-matching angle. The pulse energy is changed with an attenuator, in which a half-waveplate is mounted on a motorized rotation stage. It is possible to obtain more attenuation by setting up multiple attenuators in series or more simply by the use of neutral density filters. The pulse timing is described in Sec. \ref{sec:PulseTimings}.

\section{Laser Beam Transport}
\label{sec:LaserBeamTransport}
\begin{figure}
\includegraphics{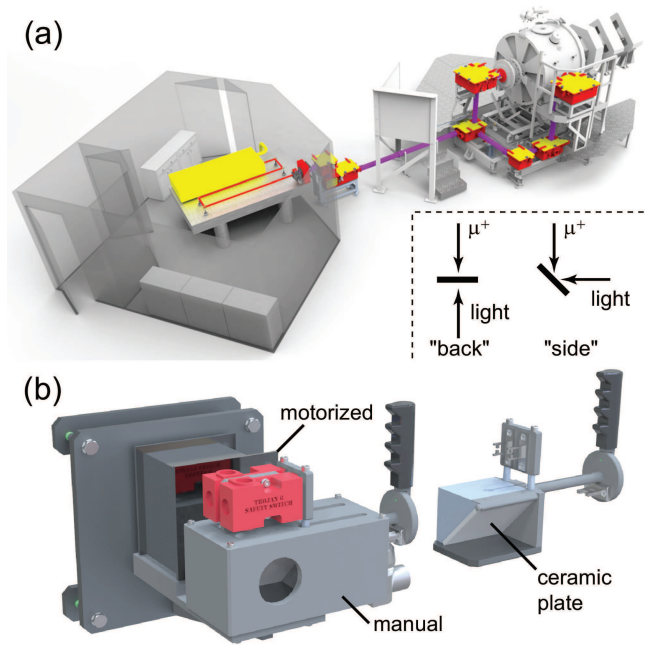}
\caption{\label{fig:FullPicture} (Color online) (a) Bird's eye view of the HiFi Laser cabin and the beam transport system. A part of the false floor and stairs in the instrument area are hidden on this figure for clarity. Inset shows the back- and side-pump geometry for the sample. (b) Left: external appearance of the beam shutter assembly, made of a manual and motorized shutter.  Right: the beam block taken out of the manual shutter. A piece of ceramic plate (Macor) is fixed on top of the triangular steel block, which is tilted at 45 degrees to increase the illuminated area.}
\end{figure}

To carry out photo-$\mu$SR experiments safely in a user-shared facility such as ISIS, the laser light needs to be contained and transported to the instrument in a fully enclosed beam transport system. The HiFi spectrometer is built around the 5T superconducting split pair magnet (and fast-sweep  z-axis coils up to 40 mT as well as 15 mT x- and y-axis transverse coils) with a room temperature internal bore along the beam axis, which contains the detectors and four transverse ports (sides, top, and bottom). \cite{Lord} These ports form a ``cruciform'' which is usually pumped and provides an insulating vacuum for cryostats. As shown in Fig.~\ref{fig:FullPicture}(a), the barrel-shaped housing containing the magnet and cruciform is used as a light-tight enclosure for the photo-$\mu$SR setup. We then need an enclosed beam transport system that connects the instrument and the laser cabin where the laser system is stored and operated.

\subsection{Laser cabin}
As shown in Fig.~\ref{fig:FullPicture}(a), the laser cabin has been built down-stream of the HiFi beamline. Since the cabin is located outside the radiation-controlled area, the laser operator can access the cabin without disturbing the on-going $\mu$SR experiment. The cabin houses a 1.5 $\times$ 2.5 meter optical table supported by rigid legs. Although the non-isolating table supports are more vulnerable to floor vibration than pneumatically isolated ones, it is more important to maintain the table position relative to the muon spectrometer by sharing the same solid floor rather than to minimize vibration. Floor vibration generated in the synchrotron magnets can have a significant impact on the optical alignment. In the current case, the vibration initially observed on the tabletop has been significantly reduced by placing the heavy laser heads and additional lead blocks on the table. As a result the floor vibration barely affects the beam alignment.

The laser cabin is equipped with interlocked doors and laser shutters for a safe laser operation. The double-door structure restricts laser light to the cabin by allowing only one door to be open at a time. Mechanical (trapped keys) and electrical interlock systems are installed on the doors, and the restricted access is enforced with a card scan system. The only way out for the laser light is through the custom-made shutter assembly, which can block the high power laser light. As shown in Fig.~\ref{fig:FullPicture}(b), the assembly is composed of a manual and motorized shutter connected in series. The manual shutter is made of a triangular block of stainless steel covered by a 10 mm thick ceramic plate. When the shutter is closed, the laser light is scattered and diffused on the tilted porous ceramic, and absorbed by the steel block. A damage test has been performed with the full-power collimated 1064-nm laser light, and after 5-hours continuous illumination, there was no visible damage on the ceramic surface when viewed under a microscope. The motorized shutter (LS-100-12 from Lasermet, Ltd.) contains a fail-safe laser blade, which opens/closes synchronously with the manual shutter, and its position is linked to the interlock system.

When the shutter assembly is open, any disturbance of the interlock system results in closure of the motorized shutter and isolation of the mains to the laser power supply. All the interlocks, including the switches on connections between tubes and vacuum flanges (see the next section), must be completed before the laser light is transported to the instrument. However, there is another mode when it is possible to operate with the laser light confined to the cabin by closing and locking the manual shutter. This is useful for the laser maintenance or development work isolated in the laser cabin, whilst the interlock switches outside the cabin are disabled and another non-laser experiment is in progress.

\subsection{Beam transport system}
The beam transport system (BTS), as shown in Fig.~\ref{fig:FullPicture}(a), is constructed from the mirror boxes, transport tubes, and beam entry chamber (BEC, the last mirror box attached on the HiFi spectrometer). Because these parts are located outside the laser cabin, all connections of the tubes and lids on the mirror boxes are controlled by the interlock system. The BEC and some of the mirror boxes have ports for electrical feedthrough, which can be used for laser diagnosis and remote control on optical components. In the instrument area, the BTS (except BEC) is hidden under the false floor so that it does not obstruct normal $\mu$SR experiments. Materials used in the area are non-magnetic because the stray magnetic field from the magnet can potentially apply a force on magnetic materials and misalign the laser beam.

Two experimental geometries are schematically shown in the inset of Fig.~\ref{fig:FullPicture}(a). When the BEC is mounted on the ``back'' of the instrument, muon and laser light counter-propagate each other and are implanted from opposite sides of the sample. In the ``side'' pump geometry, the sample may be tilted to accept both muon and laser light on the same side. The back-pump geometry is easier for the optical setup and has been used in all experiments carried out so far. In this geometry it is important to consider both the absorption length of a given wavelength and the muon implantation depth to ensure a good muon/photon spatial overlap for maximum light effect. \cite{Murahari}

The laser beam, after leaving the cabin, is routed toward the instrument by several mirrors. Each mirror box has an optical breadboard in the bottom to set up mirrors and other optical components. The beam is aligned using a 405 nm Class II diode laser (DL) set on the optical table in the cabin. The DL beam is firstly aligned to the mirrors in a safe operation mode, which disables all the interlocks but inhibits the main laser. At the sample position (in back-pump geometry), the DL beam is aligned to a target ``jig'' mounted on the cold finger of the closed cycle helium refrigerator (CCR), which indicates the center of the muon beam. The target jig has also been imaged with the low-light CCD camera used for observing the muon beam spot. \cite{Lord} Once the beam path is defined, two iris diaphragms aligned to the DL beam are positioned far apart on the optical table. The real laser beam is then aligned to the irises, and normally follows the pre-defined beam path down to the sample. The alignment sometimes requires an adjustment during the experiment because a slight change in the direction of the beam can result in a significant displacement of the beam position after propagation over the 8 m distance from the beam shutter to the sample position in HiFi.

\begin{figure}
\includegraphics{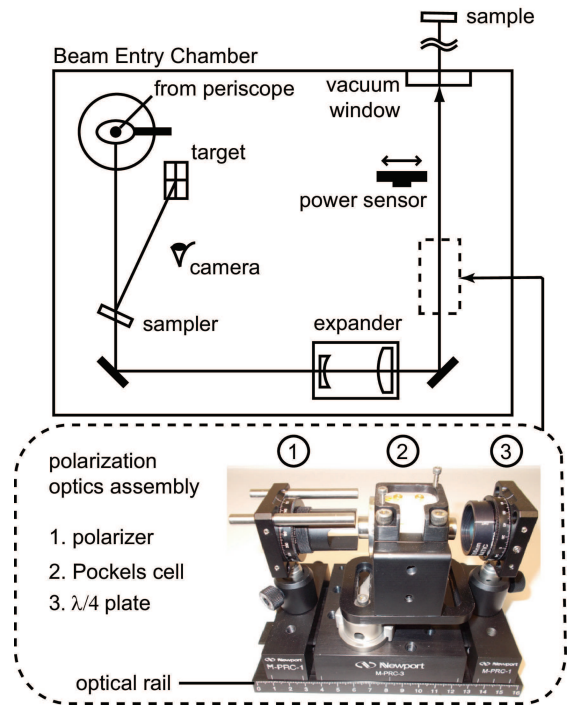}
\caption{\label{fig:BEC_BeamPath} (Color online) Schematic birds-eye view of the BEC. The polarization optics assembly is pre-aligned on the optical rail and installed in the dotted square for producing circularly polarized light. In that case the beam expander will be moved to a position after the assembly because of the small apertures of the optics.}
\end{figure}

Fig.~\ref{fig:BEC_BeamPath} shows a typical optical setup in the BEC for the back-pump geometry. The beam coming out of the periscope (see Fig.~\ref{fig:FullPicture}(a)) is reflected and leveled to the breadboard, then incident on to a beam sampler. The partial reflection aligned to a target is monitored by a camera to help adjusting the beam position. The long distance between the first iris and this target guarantees that the real laser beam illuminates the right spot on the sample. The beam size in BEC is typically $\approx$15 mm in diameter, which is then enlarged to match the size of the muon beam using an expanding telescope of 2$\times$ $\sim$ 4$\times$ magnification. If the power allows, it is ideal for the laser light to cover an entire area of the muon beam to cancel out small misalignment and any possible drift. The telescope distance is adjusted in advance in the laser cabin because the interlock system requires the BEC to be closed while the high power laser light is present inside. The beam spot in the BEC can be reproduced on the optical table by propagating the beam for the equal distance --- this is a general method for observing/modifying the beam when an in-situ access is restricted. Laser power can be measured in front of the exit port of BEC using a power sensor. The measured pulse energy is calibrated with another power sensor in the laser cabin, which measures a partial reflection from a beam sampler, therefore the pulse energy on sample is live-monitored and recorded. 

The multiple modes in the laser beam increases energy dissipation during the transport. The transmission efficiency to the BEC depends on the laser source (either YAG or OPO) and wavelength, and ranges between 30 and 70 \%. A large beam, and therefore large-sized mirrors and optics, helps to reduce the power loss caused by the long propagation distance. Thus the internal diameter of the transport tubes is 10 cm, and the BTS currently supports mirrors up to 2-inch diameter. If an experiment requires the best beam quality and the highest power transmission, an image relay telescope with long focal distance lenses can be installed between the bottom mirror boxes in the BTS. The BTS can also be pumped and purged with inert gases, such as Ar, to reduce absorption losses by oxygen and water in air, which are pronounced in the deep UV and mid-IR range respectively.

The tunable OPO laser system requires multiple sets of mirrors to cover the entire spectral range. Currently we use three mirror sets: 1) protected silver mirrors for low-power visible and IR (PF series from Thorlabs Inc.), 2) broadband dielectric mirrors for high power beam in 350 - 1100 nm range (BBDM series from Semrock), and 3) UV broadband dielectric mirrors for wavelength in 245 - 390 nm range (MAXBRIte series from CVI). Further mirror sets can be purchased as required.

\section{EXPERIMENTAL TECHNIQUES}
\label{sec:ExperimentalTechniques}
In this section we first focus on techniques that ensure that the muons and photons overlap in a sample in both time and space. It is essential to manage this overlap for a controlled photo-$\mu$SR experiment. Sec. \ref{sec:Sample} describes three types of sample environment currently available for photo-$\mu$SR experiments in HiFi. For convenience, their specifications are summarized in TABLE \ref{table:SampleEnvSpecs}. Finally Sec. \ref{sec:Polarization} illustrates how the circular polarized light is generated and manipulated in our system.

\subsection{Pump-probe pulse timings}
\label{sec:PulseTimings}

\begin{figure}
\includegraphics{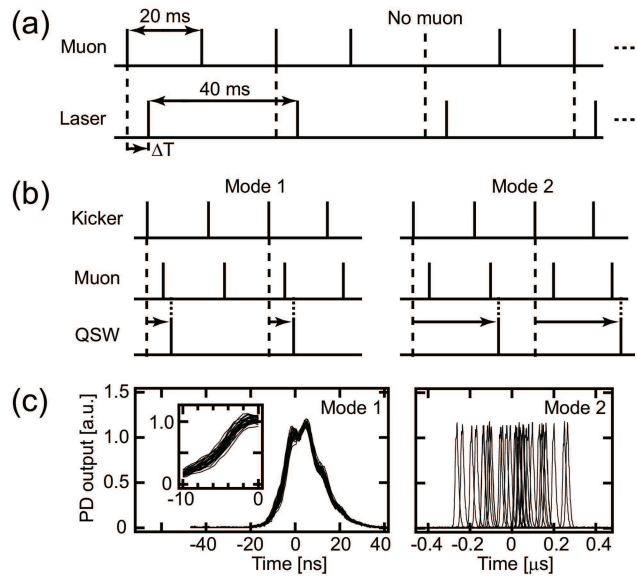}
\caption{\label{fig:PulseTimingJitter} (a) Timing diagram for muon and laser pulse. The delay generator makes an arbitrary delay, $\Delta$T, on the laser pulse. (b) Timing diagram for timing Mode 1 and 2 shows how the Kicker signal triggers the QSW to synchronize with the muon pulse. Note both diagrams illustrate the same $\Delta$T. (c) Laser pulse jitter for each mode. A total of 32 pulses have been measured using a fast photodiode and oscilloscope. Inset is a zoomed section of the rising edge of the Mode 1 waveform to show the jitter more clearly.}
\end{figure}

The laser pulse is synchronized with the muon pulse using a digital delay generator (Stanford Research Systems, DG645), which triggers the YAG flashlamp and Q-switch (QSW) with a set time delay. As shown in Fig. \ref{fig:PulseTimingJitter}(a), the ISIS Target Station 1 currently runs on a ``pseudo'' 50 Hz with a missing pulse after every 4 consecutive pulses, whereas the laser system runs at a steady 25 Hz. The delay generator sends a signal to the muon data acquisition equipment (DAE) to sort the data bins for the ``light ON'' and ``light OFF'' spectra. \cite{Giblin} By taking the difference of the spectra, (light ON) \verb|-| (light OFF), one can remove not only the systematic errors and drift but also the instrument background in the field scan. \cite{Lord}

As shown in Fig. \ref{fig:PulseTimingJitter}(b), there are two ways to trigger the laser system depending on the pulse delay, $\Delta$T (positive for laser pulse arriving after muon). The delay generator is triggered by the ``Extract Kicker'' signal that kicks a proton bunch from the synchrotron ring. Generated muons arrive at the instrument $\approx$3.5 $\mu$s after the kicker signal. Including the circuit delay in the laser system ($\approx$0.5 $\mu$s), one can set the delay for \verb|-|3.0 $\mu$s \verb|<| $\Delta$T \verb|<| 20 ms in this trigger mode (Mode 1). It is, however, not useful to have $\Delta$T \verb|>| 32 $\mu$s because the light then arrives after all the muons have already decayed. To put the laser pulse earlier than $\Delta$T = \verb|-|3.0 $\mu$s, the ``previous'' kicker signal needs to be used for the trigger. Therefore this timing Mode 2 covers the delay ranging \verb|-|20 ms \verb|<| $\Delta$T $\leq$ \verb|-|3.0 $\mu$s. The disadvantage of Mode 2 is the relatively large pulse-to-pulse jitter inherent in the operation of the proton synchrotron, where the kicker has to be phased with the circulating proton bunches. Fig. \ref{fig:PulseTimingJitter}(c) shows oscilloscope traces for 32 laser pulses in each timing mode. In Mode 1 the p-p jitter is in an order of a nanosecond, which is associated with the build-up time in the YAG oscillator. In Mode 2, on the other hand, the jitter is $\pm$300 ns p-p with a symmetric triangular distribution, which is manageable with a delay time of around 3 $\mu$s and increasingly negligible at the longer pump-probe separation times.

\subsection{T$_0$ measurement}

\begin{figure}
\includegraphics{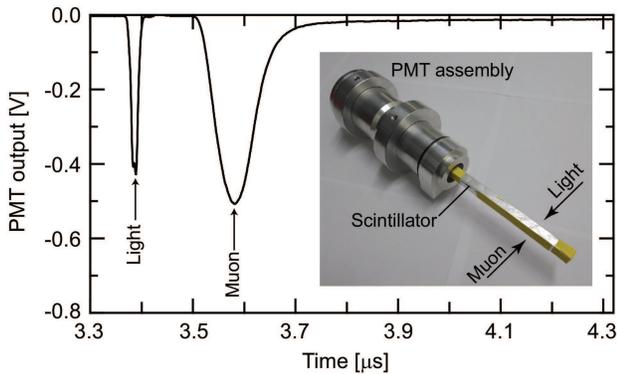}
\caption{\label{fig:T0Measurement} (Color online) Oscilloscope waveform for the T$_0$ measurement shows the muon and laser pulse (532 nm). The waveform has been averaged over 512 shots. The long tail of the muon pulse is associated with the muon-positron decay with 2.2 $\mu$s lifetime.}
\end{figure}

For all photo-$\mu$SR measurements, especially for those observing fast dynamics, it is important to define a time zero, $\Delta$T = 0 (T$_0$), when the muon and laser pulse are implanted simultaneously in a sample. This is not trivial because all electronics/cables introduce some delay in the signal transmission and change the pulse timing. Thus we have developed a method using a single detector and common electronic processing, as shown in the inset of Fig. \ref{fig:T0Measurement}. The apparatus is composed of a long rectangular polished plastic scintillator attached on to a photomultiplier tube (PMT). The scintillator column is wrapped with an aluminum (Al) foil to exclude ambient light which could saturate the PMT. The foil has a pinhole on the back side where the laser beam is incident. When the assembly is positioned in the instrument with the scintillator part exposed to muon and laser beams simultaneously, both scintillation light from implanted muon and laser light are detected by the PMT. Fig. \ref{fig:T0Measurement} shows its oscilloscope waveform, where the value of $\Delta$T can be measured (here $\Delta$T = -200 ns). This measurement cannot run with real samples in the beam position, and thus has to be performed in advance.

Since the T$_0$ timing may vary slightly depending on the initial accelerator setting (within $\pm$50 ns in ISIS), ideally  T$_0$ should be measured for every cycle. However since this method for determining T$_0$ requires access to the instrument, it may not be convenient to repeat this too often. However, one can use the muon \v{C}erenkov counter and a photodiode on the laser table as secondary standards to confirm the timing in each experiment. The \v{C}erenkov detector is positioned next to the muon target in the beam line and provides the usual start signal for muon data acquisition, and the laser pulse is measured at a fixed position on the optical table. As long as the same beam path, electronics, and cables are used every time, the T$_0$ timing should be accurately reproduced.

\subsection{Sample environment}
\label{sec:Sample}

\begin{table}
\centering
\caption{Sample environment specifications}
\label{table:SampleEnvSpecs}

\begin{tabular}{|p{33mm}|>{\centering\arraybackslash}p{17mm}|>{\centering\arraybackslash}p{15mm}|>{\centering\arraybackslash}p{17mm}|}
\hline
 & He-purged cell & He flow cryostat (Variox) & Liquid flow cell \\
\hline \hline
Max. sample size [mm] & 50.8$\phi$ $\times$ 3t & 45$\phi$ $\times$ 5t & 32$\phi$ $\times$ 3t \footnotemark[1] \\
\hline
Cell material & Al & --- & Cu \\
\hline
Temperature range [K] & $\sim$15 - 400 \footnotemark[2] & $\sim$2.0 - 300 & 283 - 303 \footnotemark[3] \\
\hline
Optical window / & Fused silica & Mylar & Fused silica \\
wavelength range [nm]  & 200 - 2000 & 400 - 2400 & 200 - 2000 \\
\hline
\end{tabular}
\footnotetext[1] {Thickness (t) can be changed discretely with a stepped window with different thicknesses.}
\footnotetext[2] {The high temperature is limited by the indium seal. This is the same for the liquid cell.}
\footnotetext[3] {The cell currently operates in 1 atmosphere with no pressure differential across the thin muon window; modifications to prevent window deflection in vacuum would allow a temperature range 243 - 400 K.}
\end{table}

\begin{figure}
\includegraphics{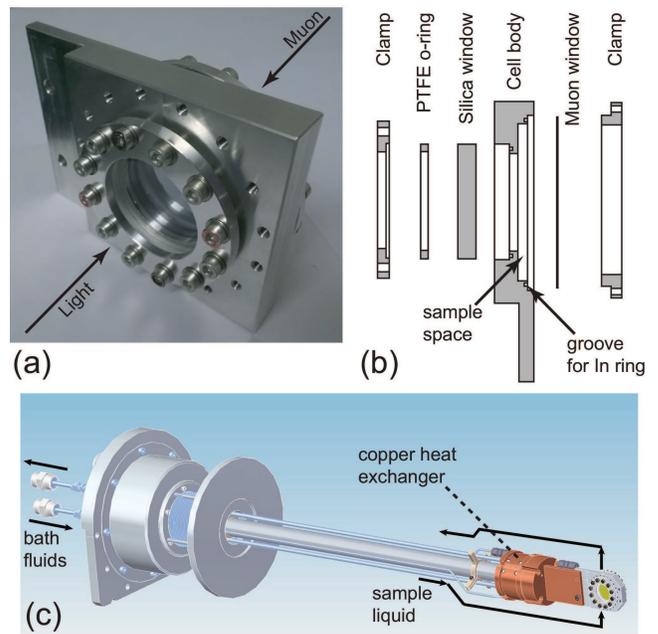}
\caption{\label{fig:Sample} (Color online) (a) Photograph of He-purged sample cell for photo-$\mu$SR experiments. (b) Cross-sectional view of the cell illustrates its structure. (c) External appearance of the liquid circulation stage and its sample cell. Arrows illustrate the flow of liquid sample.}
\end{figure}

As shown in Fig. \ref{fig:Sample}(a), a sample cell has been developed for photo-$\mu$SR experiments in the back-pump geometry. The cell is mounted on the cold finger of the CCR. \cite{Lord} A radiation shield surrounding the cell is thermally anchored to the first stage of the CCR cooler (50 K), and has apertures in the front and back faces. The aperture for the muon beam is covered with a piece of thin Al foil for a better thermal insulation. Fig. \ref{fig:Sample}(b) shows its structure in a cross-sectional view. The optical window is made of fused silica, and all other parts are made of Al for use in high fields. Samples are contained in the narrow space indicated by an arrow, which can store samples up to 50.8 mm diameter and 3 mm thick. The sample cell is assembled in a helium glove box so that the contained He gas works as thermal exchange gas. The seal is made by 1-mm diameter indium O-rings. The muon implantation depth can be adjusted by changing the number of Al foil degraders, and/or changing the thickness of the muon window.

In this cooling setup, because the thermal exchange gas equilibrates with the cell, samples can be cooled uniformly and efficiently down to the base temperature of the cryostat. It is also important to note that this cell can cool samples without mechanical stress, which can affect some of the material properties, such as electron spin lifetime in GaAs, where the lattice distortion drastically affects the lifetime. \cite{Crooker} Possible ``flopping'' of sample can be avoided with a small amount of low-temperature grease applied between the sample and cell body.

Cooling performance of the cell has been confirmed using intrinsic silicon. It is known that the diamagnetic fraction of $\mu$SR signal in intrinsic Si, as measured in a weak transverse field, decreases monotonically from 250 K down to 150 K. This behavior is attributed to slowing down of the thermally activated ionization of Mu centers into diamagnetic centers. Hence it can be used as an approximate probe of the sample temperature, and has confirmed the cooling performance of the cell. 

For temperatures below the base temperature of CCR ($\approx$10 K), a helium cryostat (2 - 300 K) loaded from the top of HiFi is available for photo-$\mu$SR applications. We fit transparent Mylar vacuum windows and make holes in the foil radiation shields just large enough to let the laser light in for sample illumination. The beam alignment in this setup requires extra care because contraction shifts the sample position as the sample-mounting stick is cooled. To adjust the laser beam spot on the sample after cooling, a piece of silicon or germanium is mounted on the stick instead of a real sample, and the beam position is scanned using a remote-controlled actuator attached to the last mirror in Fig. \ref{fig:BEC_BeamPath}. The light ON/OFF effect is easy to see in these materials, and the beam should be correctly aligned when the maximum (light ON) \verb|-| (light OFF) signal is observed. It is also important to note that the local heating from laser light can easily change the sample temperature in these low-T experiments.

As mentioned in Sec. \ref{sec:Introduction}, one of the important applications of the photo-$\mu$SR technique in high field is to study the organic molecular radicals in excited states. Muoniated molecular radicals are efficiently created in a group of liquids ({\it e.g.} water, hexane, and alcohol), \cite{Walker} where a large fraction of implanted muons are converted to solvated Mu, which then chemically attacks on electron-rich parts of the solute molecule. Typical molecular excitation lifetimes are in the ns to $\mu$s range, and given the laser pulse separation is 40 ms, the molecules repeatedly cycle through the excited and ground states. The presence of the excitation often makes the molecules more chemically reactive, such that one has to be careful with the photodegradation timescales when planning experiments. To mitigate some of the issues around photodegradation, it is possible to circulate the solution to replace the reacted molecules with new ones. Depending on the chemical reactivity and degradation mechanism of the molecules in a given experiment, it may be possible to circulate via a closed-loop or it may be necessary to dump the solution after illumination, continually refilling the cell with fresh solution. Fig. \ref{fig:Sample}(c) shows the newly developed liquid circulation stage and the liquid sample cell, which is mounted on a slider in the side of HiFi. The liquid cell has the similar structure as the He-purged sample cell in Fig. \ref{fig:Sample}(a), but has a pair of copper tubes soldered on its top and bottom. These tubes are routed out through the flange and go into a sample reservoir and a peristaltic pump. The circulation system injects the liquid into the cell from its bottom and takes it out from the top as indicated by arrows in Fig. \ref{fig:Sample}(c). The tubes and sample cell are thermally anchored on a copper heat exchanger, which is temperature-controlled by a circulating oil bath. The liquid temperature is in equilibrium with the heat exchanger as long as the flow rate is sufficiently low (typical flow rate \verb|<|1 mL/s). 

The solvent must  be deoxygenated prior to use, \cite{Walker} either by the traditional freeze-pump-thaw (FPT) method or by bubbling an inert gas, such as Ar, through it for a number of hours prior to the experiment. The former method has the advantage of that it is compatible with almost every solvent, it is clear when the solvent is deoxygenated (based on the FPT vacuum) and is ideal for small volumes, although the current system requires prolonged solvent preparation if one is deoxygenating large volumes ({\it e.g.} 0.5 L). The bubbling technique has the advantage that it can easily deoxygenate large volumes without much manual intervention, but it is incompatible with solvents with a low boiling point ({\it e.g.} dichloromethane) and it is not always clear when the solvent is deoxygenated. However, it has the considerable advantage of being able to be performed in-situ, with a glove box that can be attached to the HiFi spectrometer. The glove box is needed irrespective of the deoxygenation method used, as over the course of a several day experiment, air can dissolve back into the solvent if it is not maintained in an inert atmosphere.

Probably the most important  advantage of a dissolved sample is the controllable absorption length. Assuming the solvent is transparent to the pump light, one can adjust the absorption length by changing the solute concentration. Typical concentrations needed are in the 10 mM range, which is ideal for performing ALC measurements, as this results in absorption lengths in many solutions in the order of  100 $\mu$m, which is comparable to the width of the muon's stopping profile. Optimization of an experiment therefore involves at least three factors: the light absorption length, muon range distribution, and the signal strength. Monte Carlo simulations can simulate the muon stopping distribution, \cite{Shiroka} and an in-depth experimental investigation of the optical properties of the material are required to optimize the light induced signal. Further technical details of this approach will be published in due course. \cite{Murahari}

\subsection{Polarization control}
\label{sec:Polarization}

By virtue of the monochromatic and short-pulsed laser light, it is possible in our system to manipulate light helicity on a pulse-by-pulse basis, using the polarization optics assembly as shown in Fig. \ref{fig:BEC_BeamPath}. The device is composed of a linear polarizer, a Pockels cell (PC), and a quarter-waveplate. The light polarization is linear in the laser cabin, but becomes slightly elliptical (and tilted) in the BEC because the mirrors in BTS are not aligned perfectly at right angles, hence each mirror introduces a small phase delay between the polarization axes. Therefore the linear polarizer first ``cleans'' its polarization before the PC, which then introduces a $\lambda$/2 retardation upon receiving a trigger. The fast axis of the $\lambda$/4 waveplate is tilted 45 degrees with respect to the polarization axis as defined by the polarizer. Therefore depending on the PC state (``PC ON'' or ``PC OFF''), the incident polarization is polarized either vertical or horizontal, which results in right- or left-circularly polarized light after the $\lambda$/4 waveplate.

The PC trigger runs at 12.5 Hz and is synchronized with the QSW using the delay generator. In the muon DAE, data will be sorted in four bins: ``light OFF 1'', ``light ON, PC ON'', ``light OFF 2'', ``light ON, PC OFF''. Two ``light OFF'' bins are essentially identical and summed in the analysis. The polarization effect is observed by taking the difference, (PC ON) \verb|-| (PC OFF), which is also a robust quantity against the high field background and experimental drift.

The polarization optics on the optical rail are pre-aligned in the laser cabin and installed in the dotted square shown in Fig. \ref{fig:BEC_BeamPath}. The advantage of changing light polarization in the last arm of the beam path is to null the difference of number of photons in the  (PC ON) and (PC OFF) state. For example, if the PC is located in the laser cabin and sends P- or S-polarized light down to the BEC, their pulse energies will be different at the end because of the different reflectivity from mirrors for each polarization. Even if the difference is small, it accumulates as the beam is reflected off many mirrors in the BTS, and poses an issue especially when the polarization effect is small but the overall light induced signal is large and dependent on intensity. The disadvantage of having the polarization optics in BEC is that alignment is more difficult and so the extinction ratio may be lower. Nevertheless an extinction ratio of $\approx$100 in the ``PC ON'' state is achievable in the current system.

\section{photo-$\mu$SR experiment on silicon}
\label{sec:Silicon}
The laser system is now included in the HiFi instrument suite and open for the ISIS user program. To demonstrate the effectiveness of the newly constructed system, we have performed a photo-$\mu$SR experiment on silicon, which is known to give a large photo-induced effect. \cite{KadonoPRL, KadonoPRB, Fan}

The experiment has been carried out on a 500-$\mu$m thick intrinsic silicon wafer (n-type, R $\approx$2400 $\Omega$$\cdot$cm) with $\langle$110$\rangle$ crystal axis perpendicular to the surface. One side is polished and is facing the incoming laser light, whereas the other side has an etched surface that faces the muon beam (back-pump geometry). The muon distribution is centered at the middle of the sample by adjusting the number of Al foil degraders. In this experiment, 355-nm laser light was used to photoexcite electron-hole pairs. The pulse energy on the sample was $\approx$2 mJ, creating about 5$\times$10$^{14}$ electron-hole pairs per cm$^2$. Immediately after the illumination, generated carriers are concentrated near the incident surface because of the short absorption length, and then diffuse into the bulk, where they interact with the implanted muons. The sample was cooled to 50 K, where most of the implanted muons formed muonium either in the bond-center sites (Mu$_{BC}$$^0$) with an axial asymmetry along $\langle$111$\rangle$, or in the isotropic interstitial tetrahedral sites (Mu$_{T}$$^0$). \cite{Patterson} 

\subsection{Time spectrum}

\begin{figure}
\includegraphics{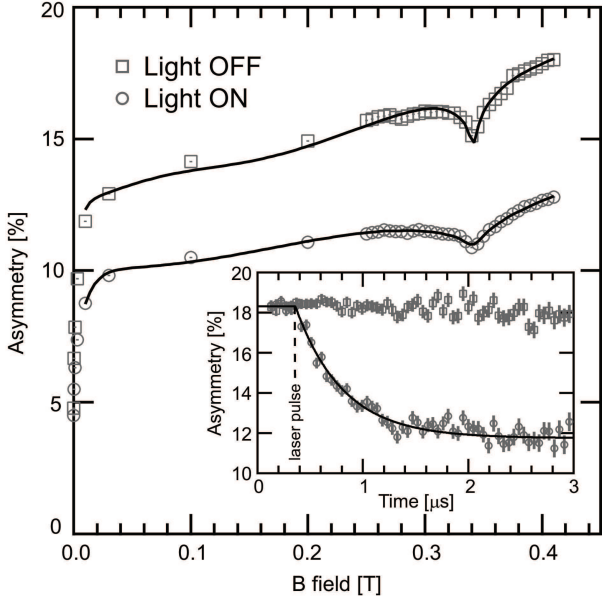}
\caption{\label{fig:Commissioning}
{\it Inset}: $\mu$SR time spectra for light OFF (open squares) and ON (open circles) from low-T (50 K) intrinsic silicon under longitudinal field (0.41 T). The $\langle$110$\rangle$ crystal axis is parallel to the field. Dashed line indicates the timing of the laser pulse ($\Delta$T = 0.36 $\mu$s). Light ON spectrum after the illumination is fitted with a single exponential with a time constant 0.45 $\pm$ 0.02 $\mu$s.
{\it Main}: Integrated muon asymmetry versus applied longitudinal field for light OFF and ON. The pulse timing is the same as inset. 
A slight misalignment between the $\langle$110$\rangle$ axis and the field gives a finite width of the dip at 0.34 T, accompanied by oscillations in the time spectra from which the alignment can be determined as $\langle$0.73, 0.68, 0.074$\rangle$. Solid lines denote the simulation results. 
For the light OFF data, the fixed parameters for fitting are: isotropic HF constant for Mu$_{T}$$^0$ = 2006.3 MHz, isotropic HF constant for Mu$_{BC}$$^0$ = -67.33 MHz, and dipolar HF constant for Mu$_{BC}$$^0$ = 50.52 MHz. Then the fitting determines the following parameters: (1) The instrument background is modeled with a quadratic function, $A x^2$, with $A=12.60$, (2) Ratio of Mu population (Mu$_{BC}$$^0$, Mu$_{T}$$^0$) = (0.62, 0.38), and (3) electron spin relaxation rates ($\lambda _{BC}$, $\lambda _{T}$) = (0.811, 0.695). Then based on these parameters, the light ON data has been fitted with a change in the electron spin relaxation rates at $\Delta$T = 0.36 $\mu$s with $A=8.31$. The relaxation rates after illumination are ($\lambda _{BC}$, $\lambda _{T}$) = (23.0, 7.95).}
\end{figure}

As shown in the inset of Fig. \ref{fig:Commissioning}, when there is no light, the muon time spectrum under LF 0.41 T shows little relaxation because the Mu HF interaction for both Mu$_{BC}$$^0$ and Mu$_{T}$$^0$ are negligibly small compared with the Zeeman splitting energy, and transitions between the Mu energy levels do not occur. The light illumination, on the other hand, induces a large relaxation in the spectrum. Generated electrons and holes interact with Mu in a quite complex mechanism including spin exchange interaction, cyclic charge exchange reaction, and site change reaction. \cite{KadonoPRL, KadonoPRB, Fan} In addition to the microscopic mechanisms, the carrier recombination lifetime and carrier transport from the illuminated surface influence the relaxation rate. Therefore, although the light-ON spectrum can be fitted well with a single exponential, the relaxation is a convolution of these factors. The amount of pulse energy deposited on the sample hardly changes the temperature: if the 2 mJ of laser light is converted to heat and uniformly distributed throughout the sample, it increases the sample temperature for $\approx$0.02 K based on the specific heat of Si at 50 K. On the other hand, the length scales that electrons and holes can diffuse within 1 $\mu$s in the current condition are estimated to be $\approx$170 and $\approx$100 $\mu$m respectively, which allow them to diffuse across the wafer and reach the region where muons are implanted. Hence the local heating is negligible and the observed effects are all due to the interaction with carriers.

\subsection{Field scan}

To demonstrate the unique combination of photo-$\mu$SR and ALC methods, Fig. \ref{fig:Commissioning} shows the integrated asymmetry as a function of the longitudinal field $B$, defined as
\[A(B)=\frac{N_f(B)-\alpha N_b(B)}{N_f(B)+\alpha N_b(B)}; \]
\[N_f(B)=\sum_{t_i=0}^{T_{max}}n_f(t_i,B), \quad N_b(B)=\sum_{t_i=0}^{T_{max}}n_b(t_i,B) \]
where $\alpha$ is a correction factor for detectors, $T_{max}$ is the last time of the spectrum, and $n_f(t_i,B)$ and $n_b(t_i,B)$ are the measured positron counts in forward and backward detectors at a given time $t_i$. When there is no light, Mu$_{BC}$$^0$ is first repolarized in low field (0.1 $\sim$ 10 mT), followed by a repolarization of Mu$_{T}$$^0$. The integrated asymmetry continues to increase up to $\approx$0.2 T. A broad but small feature at 0.15 T and the sharp dip at 0.34 T are associated with the level-crossing resonance of the Mu$_{BC}$$^0$ center, for bonds at approximately 35 and 90 degrees to the field respectively. The dip changes its width, amplitude, central field, and the asymmetric shape depending on the crystal orientation with respect to the applied field. The solid line shows a simulated curve using QUANTUM, a program to solve the time evolution of the muon spin using the density matrix method. \cite{LordQuantum} The simulation parameters are listed in the caption of Fig. \ref{fig:Commissioning}. In short, it assumes a Mu distribution constituted from 62 \% of Mu$_{BC}$$^0$ and 38 \% of Mu$_{T}$$^0$, and a finite spin relaxation for the bound electron in both Mu centers ($\lambda _{BC}$ and $\lambda _{T}$ for Mu$_{BC}$$^0$ and Mu$_{T}$$^0$ respectively). The simulation is generally in good agreement with the light OFF data.

When the laser light is ON, induced fast spin relaxation decreases the integrated asymmetry drastically. Kadono et al. attributes the predominant depolarization mechanism to the cyclic charge exchange reaction, where Mu$_{BC}$$^0$ undergoes a cycle, Mu$_{BC}$$^0$ $\leftrightarrow$ Mu$_{BC}$$^+$ + e$^-$, and points out that the charge exchange reaction is qualitatively the same as the spin exchange interaction when the charge exchange rate is significantly slower than the Mu$_{BC}$$^0$ HF constant. \cite{ KadonoPRL, KadonoPRB}  Therefore our simulation assumes a change in $\lambda _{BC}$ and $\lambda _{T}$ upon light illumination. The obtained electron spin relaxation after the illumination is clearly faster in Mu$_{BC}$$^0$ (23.0 $\mu$s$^{-1}$) than in Mu$_{T}$$^0$ (7.95 $\mu$s$^{-1}$), which suggests that Mu$_{BC}$$^0$ interacts more with the excess carriers than Mu$_{T}$$^0$. This observation may be understood as a manifestation of the fact that Mu$_{BC}$$^0$ is exposed to the carriers because of its position along the covalent bond, whereas Mu$_{T}$$^0$ is rather isolated in the tetrahedral interstitial site where the probability density of electrons and holes are much smaller.

In addition to the cyclic charge exchange reaction, Kadono et al. claim that photo-injected excess carrier triggers the site change reaction, Mu$_{T}$$^0$ $\rightarrow$ Mu$_{BC}$$^0$, via the hole capture process, Mu$_{T}$$^0$ + h$^+$ $\rightarrow$ Mu$_{BC}$$^+$, to model their data. \cite{ KadonoPRL, KadonoPRB} However as can be seen in Fig. \ref{fig:Commissioning}, a simulation with only electronic spin relaxations can simulate our data well. These two experiments cannot be simply compared because there are differences in experimental conditions. An obvious and important difference between our experiment and those of Kadono et al. is the light source: the flashlamp, which was used in their experiment contains all of the spectrum continuously from IR to UV, which have different absorption lengths. The 355-nm laser light, on the other hand, creates excess carriers at the very surface, and the carriers that Mu interact with diffuse from the surface. An ideal experiment would use a monochromatic wavelength in IR close to the band edge, which can distribute the excess carriers almost uniformly throughout the sample. This is in fact an experiment that the HiFi Laser system will be used to carry out in the immediate future, with a much better control on the light intensity (hence the excess carrier density), which is a key parameter to quantify the dynamics between excess carriers and Mu. \cite{YokoyamaSi}

Finally it is worth noting that the time spectra are available for each point in the field scan (as shown in the inset of Fig. \ref{fig:Commissioning}). Analyzing this time differential ALC data \cite{Kreitzman} with the computer simulation \cite{LordQuantum, LordTDALC} should be a powerful method in photo-$\mu$SR, where light can induce a significant change in the time spectra.

\section{conclusion}
The HiFi muon spectrometer at the ISIS pulsed neutron and muon source has been successfully upgraded and commissioned with a new high energy Nd:YAG-OPO laser system. The installation enables light-pump muon-probe experiments in high fields, which, in conjunction with the ALC method, provides a new and unique experimental environment. Needless to say, the instrument should be useful for general photo-$\mu$SR experiments. Techniques and devices have been developed to synchronize the muon and laser pulse timings accurately. These new sample environments enable one to implant both muon and photon in solid and liquid samples with an efficient temperature control. The demonstration experiment with silicon has shown the potential of this  unique combination of the photo-$\mu$SR and ALC techniques. The laser system is now in routine operation as a part of the HiFi instrument suite.

\begin{acknowledgments}
The construction of the HiFi Laser system was funded by European Research Council (Proposal No 307593 - MuSES). We wish to acknowledge the assistance of a number of ISIS technical and support staff involved in the project. We thank Dr. Katsuhiko Ishida in the RIKEN-RAL Muon Facility not only for providing helpful comments but also for kindly lending various equipment to us. We thank Mr. Geoff Gannaway and his colleagues in the workshop in Queen Mary University of London for manufacturing a number of components. KY would like to thank the spintronics collaboration \cite{YokoyamaGaAs} for various technical developments, which formed a basis of the present study.
\end{acknowledgments}



\begin{thebibliography}{10}
\bibitem{Note5th} A high conversion efficiency for the 5$^{th}$ harmonic has been achieved by mixing 532 and 355-nm laser light. This combination has a few advantages against the standard 1064/266-nm mixing: 1) less absorption of pump UV by the harmonic crystal (BBO), and 2) wavelength separation is easier using a harmonic separator.
\end{thebibliography}
\end{document}